\documentclass[twocolumn,pra]{revtex4-2}
\usepackage{amssymb}
\usepackage{amsfonts}
\usepackage{amsmath}
\usepackage{amsxtra}
\usepackage{amscd}
\usepackage{amsthm}
\usepackage{graphicx}
\usepackage{epsf}
\usepackage{bbold}
\usepackage{xcolor}
\usepackage{mathalfa}
\usepackage{eucal}
\usepackage{bm}
\usepackage{array}
\usepackage{hyperref}
\usepackage{xcolor}
\usepackage{tikz}
\hypersetup{
    colorlinks=true,
    linkcolor=blue,
    filecolor=blue,      
    urlcolor=blue,
    citecolor=blue,
    }

\begin{document}

\title{Solitons induced by an in-plane magnetic field in rhombohedral multilayer graphene}

\author{Max Tymczyszyn}
\author{Peter H. Cross}
\author{Edward McCann}
 \email{ed.mccann@lancaster.ac.uk}
\affiliation{Department of Physics, Lancaster University, Lancaster, LA1 4YB, UK}

\begin{abstract}
We model the influence of an in-plane magnetic field on the orbital motion of electrons in rhombohedral graphene multilayers.
For zero field, the low-energy band structure includes a pair of flat bands near zero energy which are localized on the surface layers of a finite thin film.
For finite field, we find that the zero-energy bands persist and that level bifurcations occur at energies determined by the component of the in-plane wave vector $q$ that is parallel to the external field.
The occurrence of level bifurcations is explained by invoking semiclassical quantization of the zero field Fermi surface of rhombohedral graphite. We find parameter regions with a single isoenergetic contour of Berry phase zero corresponding to a conventional Landau level spectrum and regions with two isoenergetic contours, each of Berry phase $\pi$, corresponding to a Dirac-like spectrum of levels.
We write down an analogous one-dimensional tight-binding model and relate the persistence of the zero-energy bands in large magnetic fields to a soliton texture supporting zero-energy states in the Su-Schrieffer-Heeger model. We show that different states contributing to the zero-energy flat bands in rhombohedral graphene multilayers in a large field, as determined by the wave vector $q$, are localized on different bulk layers of the system, not just the surfaces.
\end{abstract}

\maketitle

\section{Introduction}

Advances in the production of thin films of rhombohedral multilayer graphene (RMG)~\cite{pierucci15,henni16,henck18,latychevskaia19,yang19,geisenhof19,bouhafs21,shi20,kerelsky21,hagymasi22} recently culminated in the realization of high-quality films with up to fifty layers~\cite{shi20}.
Scanning tunneling spectroscopy~\cite{pierucci15}, magneto-Raman~\cite{henni16}, and photoemission~\cite{pierucci15,henck18} measurements have confirmed the existence of flat bands near the Fermi surface, as predicted theoretically~\cite{mcclure69,latil06,aoki07,arovas08,koshino09,zhang11,heikkila11,xiao11,slizovskiy19,garciaruiz19,muten21}. Meanwhile, the observation of phases including ferromagnetism~\cite{zhou21a} and superconductivity~\cite{zhou21b} in trilayers of rhombohedral graphene have been attributed to flat bands with strong electronic interactions.

\begin{figure}
    \centering
    \includegraphics[width=0.7\linewidth]{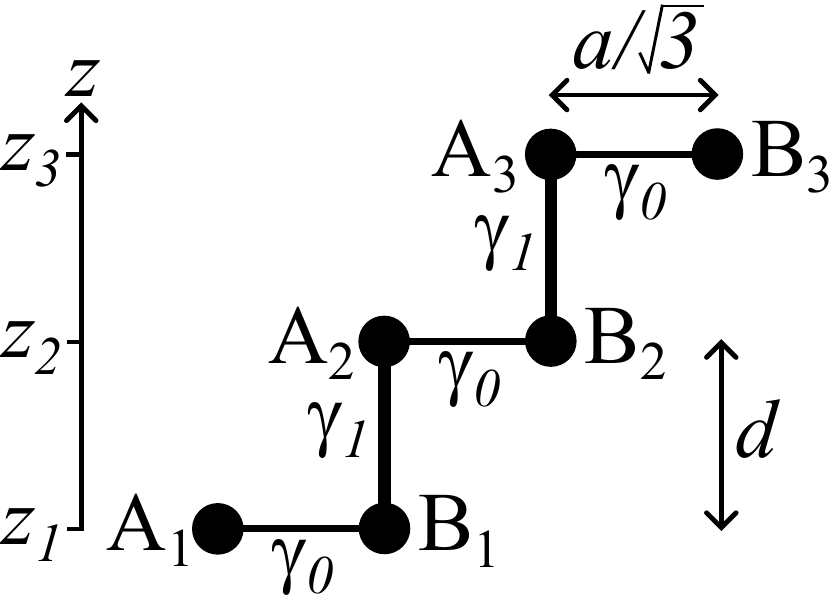}
    \caption{Schematic side view of the unit cell for three layers of rhombohedral multilayer graphene. Labels $A_n$ and $B_n$ denote the two nonequivalent atomic sites on each layer, $\gamma_{0}$ and $\gamma_{1}$ are the intralayer and interlayer hopping parameters, respectively. The in-plane carbon-carbon bond length is $a/\sqrt{3}=1.42\AA$ where $a$ is the lattice constant, while the interlayer distance $d=3.46\AA$. The out-of-plane $z$ axis is shown on the left where $z_n$ is the coordinate of the $n^{{\mathrm{th}}}$ layer.}
\label{unitcell}
\end{figure}

The presence of flat bands localized at the surfaces of RMG may be understood by analogy with edge states within the bulk band gap of the one-dimensional Su-Schrieffer-Heeger (SSH) model~\cite{su79,asboth16,cayssol21,mccann23}, whereby the intra- and interlayer couplings in RMG play the role of alternating hopping parameters in the SSH  model~\cite{xiao11,heikkila11}, Fig.~\ref{unitcell}. It has been predicted~\cite{slizovskiy19,henni16} that the bulk band gap of RMG closes with layer number $N$ at wave vector $q_{c}=\gamma_{1}/\hbar v$ near the K-points at the corner of the Brillouin zone, where $\gamma_{1}$ is the interlayer hopping parameter and $v$ is the intralayer velocity. In the limit of large $N$, i.e., bulk rhombohedral graphite, the band gap is closed and zero energy states occupy a `Dirac spiral', which rotates, as a function of the perpendicular wave vector $k_{z}$ \cite{heikkila11b,ho14,ho16,chen19}, around a K-point at in-plane radius $q_{c}$.

The study of Landau level (LL) spectra and the integer quantum Hall effect (QHE) in magnetic fields are key characterisation tools of graphene~\cite{novoselov05,zhang05} and related nanomaterials.
For out-of-plane (perpendicular) magnetic field of magnitude as little as $B=1\,$T,  Shubnikov–de Haas oscillations emerge in RMG~\cite{shi20}.
For slightly stronger fields, $B>3$T, discrete LL form~\cite{ho11,ho12,slizovskiy19}, leading to quantized Hall resistivity and the onset of the QHE observed experimentally~\cite{shi20}.
In-plane (parallel) magnetic fields have also been explored in graphene-related systems \cite{lundeberg10,vanderdonck16,slizovskiy19,qin21}, specifically in relation to the magnetic ratchet effect in bilayer graphene~\cite{kheirabadi16,kheirabadi18}, superconductivity in trilayers~\cite{zhou21b,szabo22}, and the energy spectrum of Bernal stacked multilayers~\cite{pershoguba10,goncharuk12}.

\begin{figure}
  \hypertarget{MAIN2}{}
  \includegraphics[width=\linewidth]{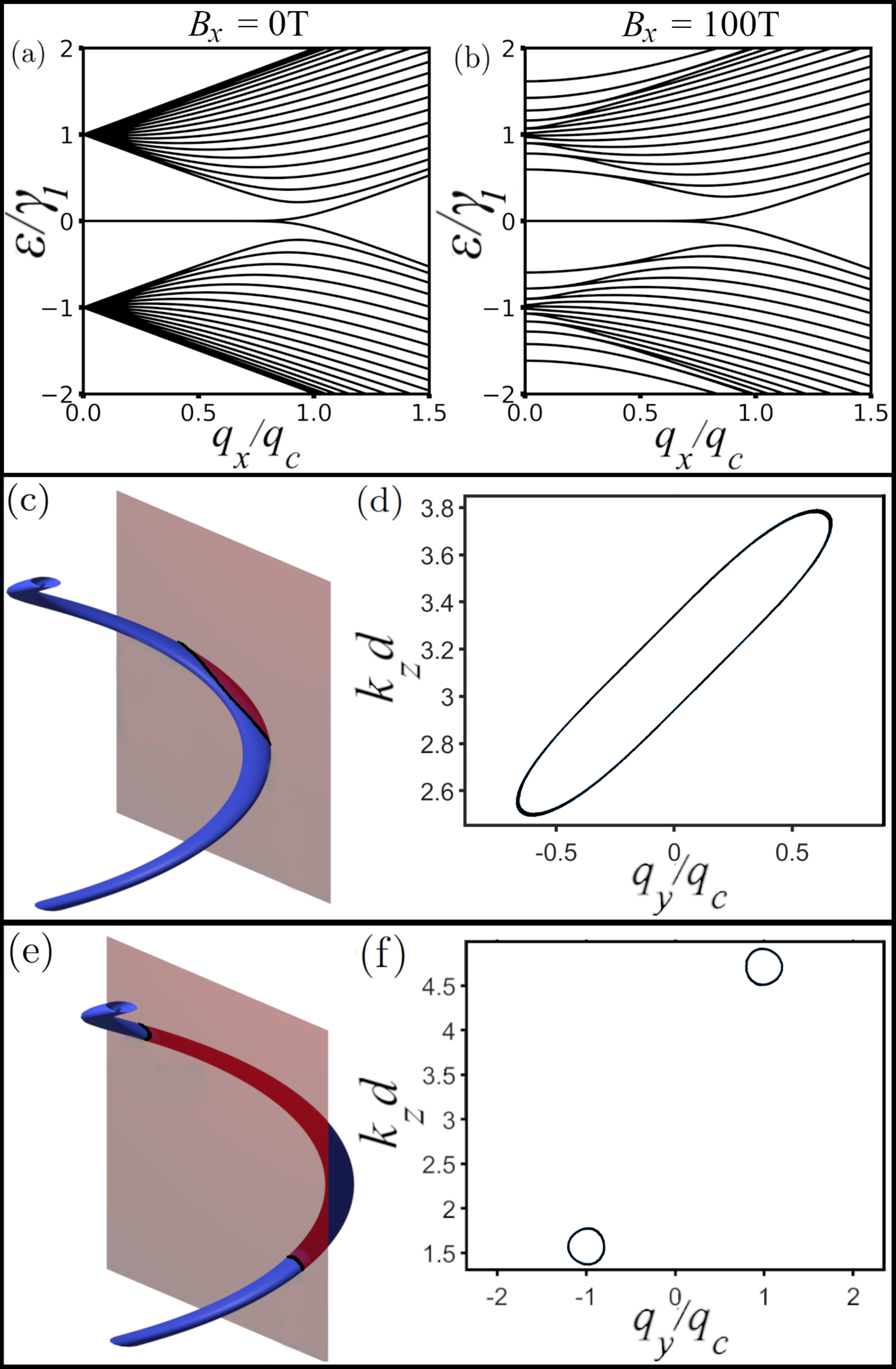}
  \caption{Low-energy band structures for rhombohedral graphene with 20 layers, with a visualisation of the corresponding semiclassical quantization showing isoenergetic contours in bulk rhombohedral graphite in a plane perpendicular to the magnetic field. (a) and (b) show band structures for no magnetic field and $B=100$T respectively~\cite{magnitudesnote}, they have been calculated through the numerical diagonalization of the Hamiltonian (\ref{HAMMY}). (c) shows an isoenergetic tube (blue) centred around the Dirac nodal spiral, a plane (red) corresponding to $q_{x}=q_{c}=\gamma_{1}/\hbar v$ intercepts the spiral (black), creating an isoenergetic contour which is depicted in (d), indicating a singly degenerate state (4-fold including spin and valley degeneracy). Figures (e) and (f) are similar, except the plane now cuts through $q_{x}=0$, creating two contours corresponding to a doubly degenerate state (8-fold including spin and valley degeneracy) found at low energy for $q_{x}<q_{c}$ in (b).}
  \label{MAIN}
\end{figure}

In this paper, we model the influence of an in-plane magnetic field on the orbital properties of electrons in RMG stacks with layer number $N\gg 1$, using 
a tight binding model~\cite{mcclure69,koshino09,slizovskiy19} and magnetic field incorporated via a Peierls substitution~\cite{kheirabadi16}. We numerically study the effect of the magnetic field on the band structure in the vicinity of the Dirac points.
Fig.~\hyperlink{MAIN2}{\ref{MAIN}(a)} shows the band structure for zero field exhibiting two flat bands at zero energy~\cite{mcclure69,latil06,aoki07,arovas08,koshino09,zhang11,heikkila11,xiao11,slizovskiy19,muten21}, and Fig.~\hyperlink{MAIN2}{\ref{MAIN}(b)} shows the spectrum for large magnetic field where we plot energy levels as a function of $q_x$ which is the component of the wave vector measured from the K point in the direction of the applied field, $B_x$~\cite{magnitudesnote}.
We note two qualitative features of the spectrum in an in-plane field: Firstly, there are a series of bifurcations from doubly- to singly-degenerate energy levels and, secondly, the zero-energy flat bands, observed at zero field, persist up to very high field strengths.

For energies $\varepsilon$, we find that the bifurcations occur along a set of points in the $\varepsilon$-$q_x$ plot, Fig.~\hyperlink{MAIN2}{\ref{MAIN}(b)}, given by
\begin{eqnarray}
|\varepsilon| \approx \hbar v (q_c - |q_x| ) , \label{beq}
\end{eqnarray}
for $|q_x| < q_c$.
The origin and position of the bifurcations~(\ref{beq}) may be understood by analysis of a two band model for bulk graphene in zero magnetic field, in which zero energy states exist along the Dirac spiral in reciprocal space near each K point~\cite{ho14,ho16}.
With an approximately linear dispersion near zero energy, an isoenergetic surface forms a tube-like structure around the spiral, Fig.~\hyperlink{MAIN2}{\ref{MAIN}(c)}.
Semiclassical quantization of LL~\cite{onsager52,kormanyos08,fuchs10,xiao10} involves quantization of the area of an isoenergetic surface in a plane perpendicular to the applied field. For field with component $B_x$, we consider cuts of the tube-like structure in the $y$-$z$ plane, where we find either one contour, Fig.~\hyperlink{MAIN2}{\ref{MAIN}(d)}, or two contours, Fig.~\hyperlink{MAIN2}{\ref{MAIN}(f)}, depending on the $q_{x}$ value.
We calculate the Berry phase~\cite{berry84} of these contours to be either zero or $\pi$ (modulo $2\pi$) for one or two contours, respectively.
Thus, the region with one contour corresponds to a `conventional' LL spectrum with constant level spacing $\sim l$ (where $l$ is the level index) whereas the region with two contours corresponds to a Dirac-like LL spectrum with doubly-degenerate levels including zero energy and level spacing $\sim \sqrt{l}$. Analysis of where the zero field spectrum transitions from one to two contours leads to Eq.~(\ref{beq}), as we discuss in further detail in Section~\hyperlink{IIb}{\ref{s:qualitative}}.

In the presence of an applied field, the spectrum retains two flat bands near zero energy. Just as these bands at zero field may be related to edge states of the SSH model~\cite{su79,xiao11,heikkila11,asboth16,cayssol21,mccann23}, then their presence in a finite field may be understood by analogy to zero-energy states localized on solitons in the SSH model as in the Jackiw-Rebbi mechanism~\cite{jackiw76,jackiw12,scollon20,cayssol21,allen22}, as explained in Section~\hyperlink{IV}{IV}.
As the magnetic field strength increases, the position of the soliton tends to move towards the center of the sample until, for very large magnetic field strengths, the solitons are annihilated and the energy levels separate away from zero energy.  Similar bound states can also be found in other systems, such as SSH models undergoing dynamic quenching \cite{rossi22} or periodic driving \cite{fedorova19}. Additionally, photonic SSH lattices have been shown to host a family of stable solitons once nonlinearity is introduced to the lattice, despite the breakdown of the bulk-boundary correspondence \cite{leykam16,gorlach17,guo20,ma21}.

Section~\hyperlink{II}{\ref{s:method}} describes the methodology beginning with the minimal model in which only nearest-neighbour hopping parameters are considered,
and Section~\hyperlink{IIb}{\ref{s:qualitative}} discusses the qualitative interpretation of level bifurcations in terms of the semiclassical quantization of the zero field Fermi surface.
Beyond the minimal model, Section~\hyperlink{IIc}{\ref{s:longrange}}, we find that the introduction of next-nearest neighbour and next-nearest layer hopping parameters does not materially affect Eq.~(\ref{beq}). An alternative calculation is presented in Sec.~\hyperlink{III}{\ref{s:infinite}} whereby we numerically determine the energy spectrum for bulk rhombohedral graphite (i.e., an infinite number of layers $N$) using a magnetic supercell. These calculations recover the results presented in Fig.~\hyperlink{MAIN2}{\ref{MAIN}(b)} for a stack with a finite number of layers $N$.
Section~\hyperlink{IV}{\ref{s:ssh}} explores the topological nature of the band structure through a comparison to the SSH model,
relating the zero-energy flat bands in a finite field to zero-energy states localized on solitons~\cite{jackiw76,jackiw12,scollon20,cayssol21,allen22}.
We also consider the influence of disorder on the spectrum by calculating the disorder-averaged density of states (DOS) per unit energy, and we find that chirality-preserving disorder does not affect the DOS in a magnetic field to a significant degree, whereas the zero-energy flat band is shown to be susceptible to chirality-destroying disorder~\cite{scollon20,muten21,allen22}.

\section{Methodology}\label{s:method}\hypertarget{II}{}

\subsection{Finite rhombohedral graphene layers with an in-plane magnetic field}\label{s:finitenumerics}
\hypertarget{IIa}{}

For numerical determination of the energy spectrum, we employ a tight-binding model~\cite{koshino09,slizovskiy19},
initially using the minimal model including nearest-neighbour intra- and inter-layer hopping only. We assume translational invariance within each layer in order to Fourier transform to $\mathbf{k}_{\vert\vert} = (k_x , k_y)$ space and create a $2N \times 2N$ Hamiltonian for $N$ layers and two atomic sites (A and B) per layer. We modify the in-plane terms $H_{AB}$ of the Hamiltonian with a Peierls substitution, which introduces a path integral of the vector potential such that \small
\begin{eqnarray*}
H_{AB}(\mathbf{k}_{\vert\vert}) \!=\! -\gamma_{0}\sum^{3}_{j=1}\exp\! \Bigg(\!i\mathbf{k}_{\vert\vert}\cdot\big(\mathbf{R}_{Bj}-\mathbf{R}_{A}\big)\!-\!\frac{ie}{\hbar}\int^{\mathbf{R_{A}}}_{\mathbf{R}_{Bj}}\mathbf{A}\cdot\mathbf{dl}\!\Bigg) ,
\end{eqnarray*}
\normalsize where $\gamma_{0}$ is the tight-binding parameter for intralayer hopping,
$\mathbf{R}_{A}$ is the position of an A atom, and $\mathbf{R}_{Bj}$ denotes the positions of three adjacent B atoms.
The magnetic vector potential $\mathbf{A}=z(B_{y},-B_{x},0)$ is chosen so that translational invariance in the in-plane direction is preserved~\cite{kheirabadi16}, and $z$ is the coordinate perpendicular to the graphene layers. Expanding and simplifying the above yields 
\begin{eqnarray*}
&& H_{AB}(\mathbf{k}_{\vert\vert}) = -\gamma_{0}\Bigg[\!\exp\!\Bigg(\!\frac{ik_{y}a}{\sqrt{3}}-\frac{iezaB_{x}}{2\hbar}\!\Bigg)\!\hspace{2.4cm}\\
&& \quad \qquad + \, 2\exp\!\Bigg(\!\frac{-ik_{y}a}{2\sqrt{3}}+\frac{iezaB_{x}}{4\sqrt{3}\hbar}\!\Bigg)\!\cos\!\Bigg(\!\frac{k_{x}a}{2}+\frac{ezaB_{y}}{4\hbar}\!\Bigg)\!\Bigg] \! .
\end{eqnarray*}

\begin{figure}
\hypertarget{largeN}{}
    \centering
    \includegraphics[width=\linewidth]{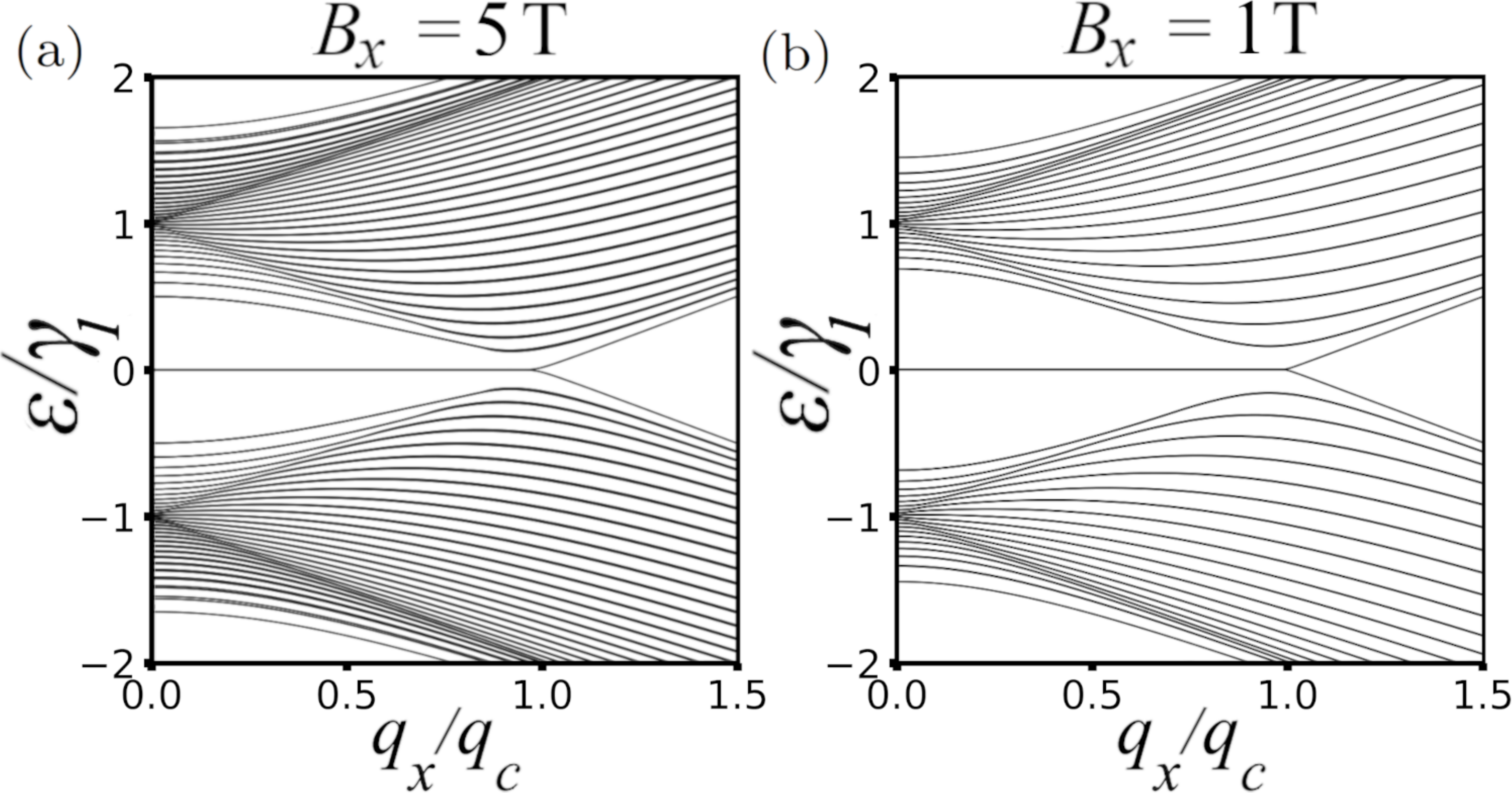}
    \caption{Low-energy band structures for rhombohedral multilayer graphene with an applied in-plane magnetic field, obtained through the numerical diagonalization of Hamiltonian~(\ref{HAMMY}). (a) $N=300$ layers and $B_{x}=5$T, where, for clearer presentation, only every $5$th pair of bands are depicted. (b) $N=1000$ layers with $B_{x}=1$T, where only every $25$th pair of bands are depicted. Bifurcations are still present at locations described by relation (\ref{beq}), but are less visible when compared to Fig.~\protect\hyperlink{MAIN2}{\ref{MAIN}(b)} due to the larger layer number.}
    \label{LargeNBands}
\end{figure}

We shift the in-plane wave vector $\mathbf{k}_{\vert\vert}$ as $\mathbf{q}= \hbar(\mathbf{k}_{\vert\vert}-\mathbf{K_{\xi}})$ to be measured with respect to the K-point at $\mathbf{K_{\xi}}=\xi(4\pi/3a,0)$ where $\xi=\pm1$ is a valley index. Then, we expand the Hamiltonian for small $|\mathbf{q}|$ and small $|\mathbf{B}|$, which is valid for $|\mathbf{q}|a / \hbar\ll 1$ and $e|z|a|\mathbf{B}|/\hbar\ll 1$, conditions which hold at low energy.
The resulting Hamiltonian reads
\begin{eqnarray}
\!\!\!\!\!\!\!\!\! H^{N}(\mathbf{q}) \!=\! 
\begin{pmatrix}
0 & \hbar v\pi_{1}^{\dagger} & 0 & 0 & \cdots & 0 & 0 \\
\hbar v\pi_{1} & 0 & \gamma_{1} & 0 & \cdots&0 &0 \\
0 & \gamma_{1} & 0 & \hbar v\pi_{2}^{\dagger}& \cdots&0 &0\\
0 & 0 & \hbar v\pi_{2} & 0 & \cdots&0 &0\\
\vdots & \vdots &\vdots &\vdots & \vdots&\vdots &\vdots\\
0&0&0&0&\cdots& 0 & \hbar v\pi_{N}^{\dagger}\\
0&0&0&0&\cdots& \hbar v\pi_{N} & 0
\end{pmatrix} \!\!\! , \label{HAMMY}
\end{eqnarray}
where $v=\sqrt{3}\gamma_{0}a / 2\hbar$ and 
\begin{eqnarray*}
    \pi_{n}(\mathbf{q}) = \xi (q_{x}+ez_{n}B_{y}/\hbar)+i(q_{y}-ez_{n}B_{x}/\hbar)\, ,
\end{eqnarray*}
where $z_{n}$ is the position of the $n^{\mathrm{th}}$ layer on the $z$ axis, Fig.~\ref{unitcell}.
We choose $z=0$ to be in the middle of the layered system. In polar coordinates we can write 
\begin{eqnarray*}
    \pi_{n}(\mathbf{q})\! = \!\!\sqrt{\!(\!q_{x}\!+\!ez_{n}B_{y}/\!\hbar)^{2}\!\!+\!\!(q_{y}\!\!-\!ez_{n}B_{x}/\!\hbar)^{2}}\,e^{i\xi\varphi_{n}}\!\!=\!  r_{n}e^{i\xi\varphi_{n}} ,
\end{eqnarray*}
and it is clear the phase $\varphi_{n}$ can be gauged away (e.g., by a redefinition of the atomic orbital wave functions). However, the magnetic field retains its influence through the magnitude $r_{n}$, and can not be gauged away.
Without loss of generality, we set $B_{y}=0$ such that the magnetic field $\mathbf{B}=\left( B_{x},0,0\right)$. By diagonalizing the Hamiltonian~(\ref{HAMMY}) we obtain the band structure of Fig.~\hyperlink{MAIN2}{\ref{MAIN}(b)}, which, when compared to the zero field band structure of Fig.~\hyperlink{MAIN2}{\ref{MAIN}(a)}, shows a spreading of the energy levels along the energy axis and the appearance of bifurcations of levels near the K-point, from doubly degenerate (8-fold including spin and valley degeneracy) to singly degenerate states (4-fold including spin and valley degeneracy). We see similar qualitative features for larger layer numbers and smaller fields, Fig~\hyperlink{largeN}{\ref{LargeNBands}}, although the bifurcations are less visible for these parameter values.
For this figure and other numerics in this paper the following hopping parameter values have been used: $\gamma_{0}=3.16eV$, $\gamma_{1}=0.381eV$, $\gamma_{2}=-0.017eV$, $\gamma_{3}=0.38eV$, $\gamma_{4}=0.14eV$ ~\cite{kuz09,bao17}, $a=2.46$\AA\; for the in-plane lattice constant, and $d=3.46$\AA\;for the interlayer spacing~\cite{bernal24}.

The qualitative features of Fig.~\hyperlink{MAIN2}{\ref{MAIN}(b)} are still observable even at comparatively low field strengths~\cite{magnitudesnote}, 
although, at $q_{x}=0$, the energy levels are only slightly spread from the zero-field degeneracy point of bulk eigenvalues at $\varepsilon = \pm \gamma_1$. At low field, the bifurcations are also localized in this region of the $\varepsilon$-$q_x$ plot, but still obey relationship~(\ref{beq}). This also holds true for small layer numbers $N\leq 10$ and larger fields, although the zero energy states do not extend to the band gap closing point $q_{x}=q_{c}$. Additionally, bifurcations are not present for $N\leq 4$ due to there being an insufficient number of bands to realise doubly degenerate states for $\varepsilon < \gamma_{1}$.
The bifurcations occur at positions~(\ref{beq}) that can be understood through a comparison to an infinite layer graphene model, as described in the next subsection.

\subsection{Semiclassical quantization of the zero-field Fermi surface}
\label{s:qualitative}
\hypertarget{IIb}{}

\begin{figure}
  \hypertarget{spiralb}{}
  \includegraphics[width=\linewidth]{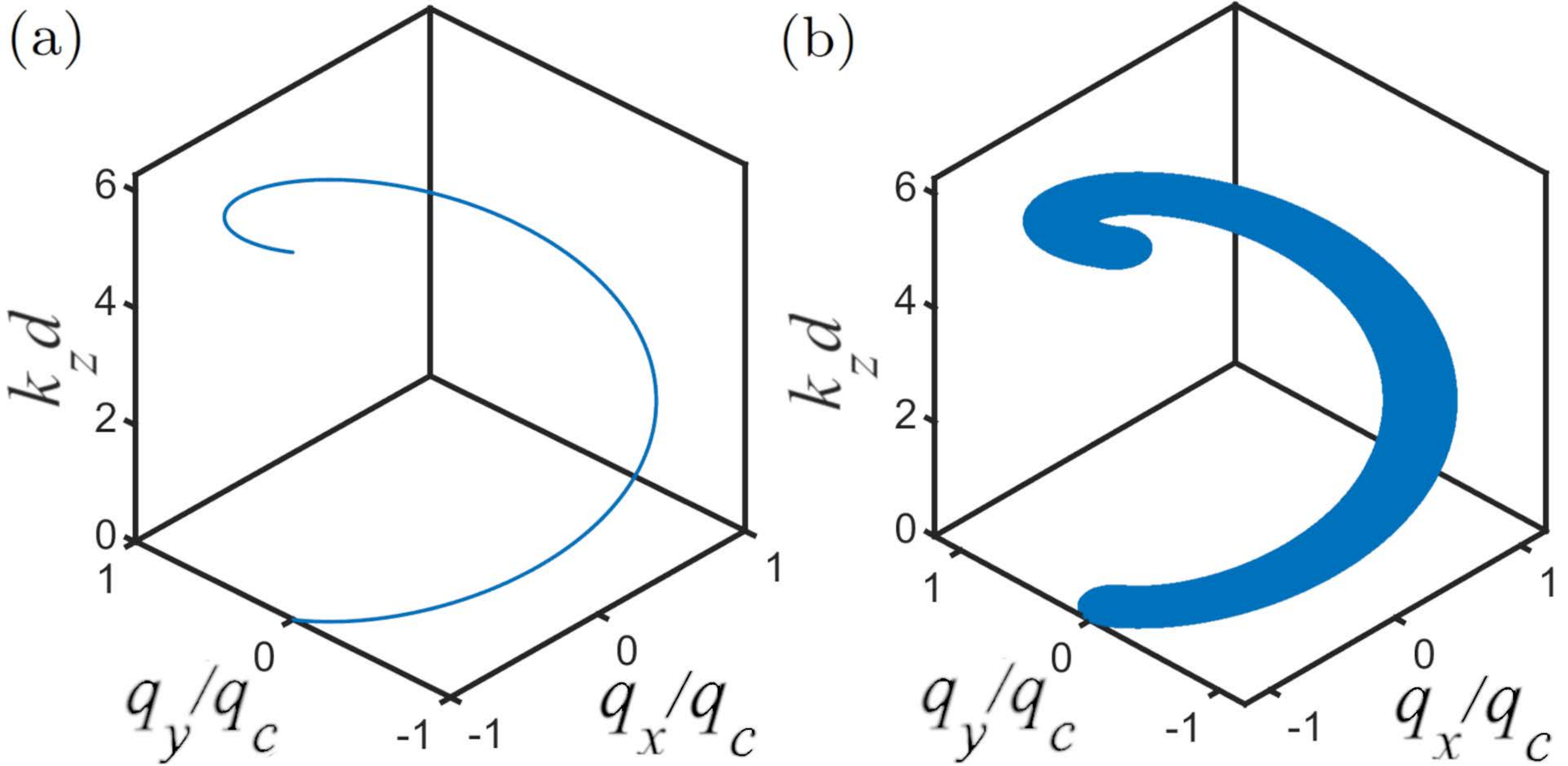}
  \caption{Dirac spirals of Eq.~(\ref{sol}) in ${\bf k}$-space located at the bulk rhombohedral graphite valley corresponding to $\xi=-1$. (a) shows the zero energy Dirac points forming a nodal line, and (b) the points of an arbitrary small energy $\varepsilon<\gamma_{1}$ centerd around zero energy, forming a `squashed' tube-like structure.}
  \label{spirals}
\end{figure}

To develop a qualitative understanding of the numerically-obtained spectrum, we consider bulk rhombohedral graphite for zero magnetic field. Assuming translational invariance in both the intra- and inter-layer directions, we introduce perpendicular wave vector $k_z$ in addition to the small in-plane wave vector $\mathbf{q}$ measured from the K-point.
By doing so the Hamiltonian \cite{ho13,ho14,ho16} is reduced to a $2\times 2$ matrix,
\begin{eqnarray}
    H^{\mathrm{inf}}(\mathbf{q},k_{z}) =
    \begin{pmatrix}
        0 & \hbar v\pi^{\dagger}-\gamma_{1} e^{i k_{z}d}\\
        \hbar v\pi-\gamma_{1} e^{-i k_{z}d} & 0
    \end{pmatrix}, \label{hinf}
\end{eqnarray}
where $d$ is the interlayer spacing and $\pi=\xi q_{x}+iq_{y}$.
In-plane wave vectors at which the energy is zero are represented in polar coordinates as $\mathbf{q}_{D} = (q_D \cos \varphi_D , q_D \sin \varphi_D )$ where $q_D = |\mathbf{q}_{D}|$ with
\begin{eqnarray*}
    q_D = q_{c} = \frac{\gamma_{1}}{v\hbar}\,,
\end{eqnarray*}
\begin{eqnarray}
    \varphi_{D} = -\xi\left(k_{z}d-\frac{\pi}{2}\right)-\frac{\pi}{2} \, .
    \label{sol}
\end{eqnarray}
The dependence of $\varphi_{D}$ on $k_{z}$ shows that these solutions represent a spiral through ${\bf k}$-space, shown in Fig.~\hyperlink{spiralb}{\ref{spirals}(a)}. The energy dispersion very close to the Dirac spiral is linear as $\varepsilon=\hbar v |\mathbf{\kappa}|$, where $\mathbf{\kappa}=(\kappa_{x},\kappa_{y})$ is the wave vector measured from the spiral for a given $k_{z}$ value.
Isoenergetic contours are circles of constant radius $\mathbf{\kappa}$ centerd around the spiral for each $k_z$ value, producing the tubular structure shown in Fig.~\hyperlink{spiralb}{\ref{spirals}(b)}. 

Onsager semiclassical quantization of LL~\cite{onsager52,kormanyos08,fuchs10,xiao10} involves quantization of the area $S(\varepsilon)$ enclosed by isoenergetic contours in ${\bf k}$-space in a plane perpendicular to the magnetic field, and is dependent on the Berry phase $\gamma$ of these contours,
\begin{eqnarray}
    S(\varepsilon) = \frac{2\pi}{\lambda_{B}^{2}}\left(l+\frac{1}{2}-\frac{\gamma}{2\pi}\right) ,
    \label{Onsager}
\end{eqnarray}
where $\lambda_{B}=\sqrt{\hbar/eB_{x}}$ is the magnetic length, and integer $l\geq0$. For example, a `conventional' 2D semiconductor has quadratic dispersion $\varepsilon = \hbar^{2}k^{2}/(2m_{*})$, effective mass $m_{*}$, and zero Berry phase. With area $S(\varepsilon)=\pi k^{2}=2\pi m_{*}\varepsilon/\hbar$, Eq.~(\ref{Onsager}) gives the harmonic oscillator relation $\varepsilon_{l} = \hbar \omega_{c}(l+1/2)$ with cyclotron frequency $\omega_{c}  = eB/m_{*}$.
By way of contrast, monolayer graphene has a linear dispersion relation $\varepsilon = \pm \hbar v k$, Berry phase $\pi$, and area $S(\varepsilon) = \pi k^{2}=\pi\varepsilon^{2}/(\hbar v)^{2}$. The Berry phase cancels the factor of $1/2$ in Eq.~(\ref{Onsager}), producing $\varepsilon_{l,\pm} = \pm(\hbar v /\lambda_{B})\sqrt{2l}$, which does not have zero point energy $\hbar\omega_{c}/2$, but a level fixed at $\varepsilon_{0}=0$.
We assume that $l=0$ is admissible in Eq.~(\ref{Onsager}), although it may relate to contours enclosing zero area, and is therefore somewhat ill-defined~\cite{kormanyos08,fuchs10}.

In this paper, we apply the semiclassical quantization Eq.~(\ref{Onsager}) to infinite layer graphite, Eq.~(\ref{hinf}), in an in-plane field in the $x$ direction. Perpendicular fields have already been studied, where the relevant contours are circles of radius $|\kappa|$ in the $x$-$y$ plane, resulting in a LL spectrum that it essentially the same as monolayer graphene~\cite{ho13}. For an in-plane field in the $x$ direction, the contours instead lie in the $y$-$z$ plane, Fig.~\hyperlink{MAIN2}{2}. The number of contours varies for different values of $q_{x}$, and we find
\begin{eqnarray*}
    \begin{cases}
        \text{0 contours} & \text{if }|q_{x}|>q_{c}+|\varepsilon|/(\hbar v)\\
        \text{1 contour} & \text {if }q_{c}-|\varepsilon|/(\hbar v)<|q_{x}|<q_{c}+|\varepsilon|/(\hbar v)\\
        \text{2 contours} & \text{if }|q_{x}|<q_{c}-|\varepsilon|/(\hbar v)\\
    \end{cases} .
\end{eqnarray*}
Moreover, when there are two contours, they have the same area.
Thus, one and two contours correspond to singly and doubly degenerate energy levels, respectively, i.e., the same degeneracies observed in the energy spectrum under an applied in-plane magnetic field.
The location at which the level bifurcations occur in the finite field $\varepsilon$-$q_x$ plot are the same as the points of transition between one and two contours in the zero field quantization.

\begin{figure}
    \hypertarget{overlays}{}
    \centering
    \includegraphics[width=\linewidth]{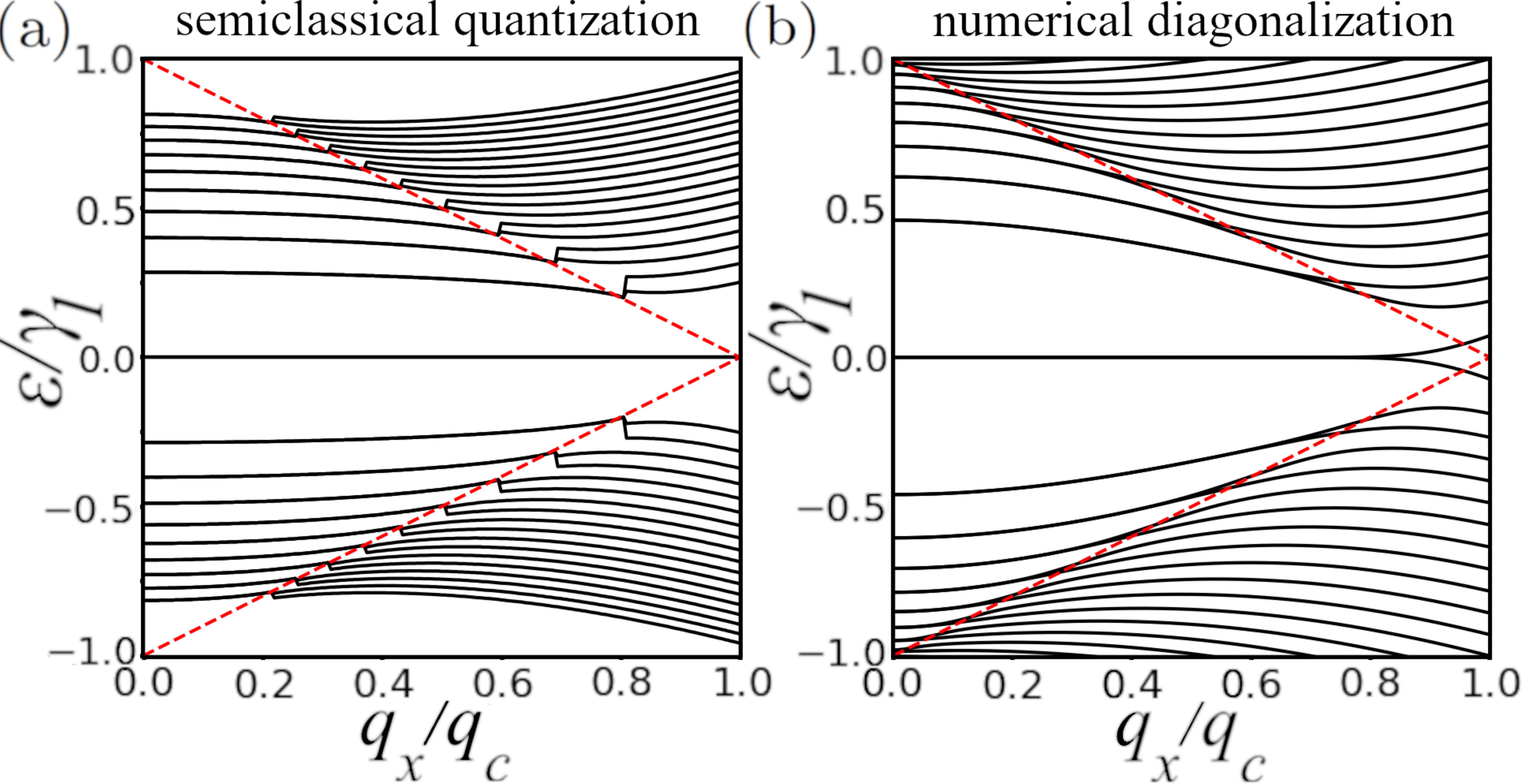}
    \caption{Low-energy band structures for rhombohedral graphene with an applied in-plane magnetic field, with an overlay (red) of the linear relationship for the location of the bifurcation points in the LL spectrum, Eq.~(\ref{beq}). (a) shows the band structure obtained from the semiclassical quantization, Eq.~(\ref{Onsager}), of the numerically obtained areas enclosed by isoenergetic contours in the infinite bulk graphite model, Fig.~\protect\hyperlink{MAIN2}{\ref{MAIN}}. (b) depicts a similar band structure for a finite 40 layer rhombohedral graphene stack with an applied in-plane magnetic field of $B_{x}=50$T, obtained from the numerical diagonalization of Hamiltonian~(\ref{HAMMY}). }
\label{overlaysplot}
\end{figure}

We find that a path around a single contour acquires a Berry phase of zero whereas the double contours each correspond to Berry phase $\pi$ (modulo $2\pi$). The latter case corresponds to a Dirac-like LL spectrum as observed in the finite field band structure at $q_{x}=0$, Fig. \hyperlink{MAIN2}{2(b)}.
At zero field and $q_{x}=0$, each of the double contours has a dispersion at low energy given by
\begin{eqnarray}
    \varepsilon \approx \pm \hbar v\sqrt{\kappa_{y}^{2}+(q_{c}d\kappa_{z})^{2}}
    \label{energyrelation}
\end{eqnarray}
where $\kappa_{z}$ is the wave vector in the $z$ direction measured from the center of the Dirac spiral.
Contours described by Eq.~(\ref{energyrelation}) are approximately circular with radius $\varepsilon$ and the corresponding enclosed area is $S(\varepsilon)=\pi\varepsilon^{2}$.
Substituting this into Eq.~(\ref{Onsager}) returns the Dirac-like LL spacing $\sim \sqrt{l}$, and, as there are two identical contours, these levels are doubly degenerate. We can extend this to finite values of $q_{x}$ by numerically calculating the contour areas, resulting in LL corresponding to an infinite system, Fig. \hyperlink{overlays}{\ref{overlaysplot}(a)}, while maintaining the same qualitative features as the band structure obtained numerically from the model of a system with a finite number of layers, Fig. \hyperlink{overlays}{\ref{overlaysplot}(b)}.
The location of bifurcations of contours in the zero field spectrum can be written as Eq.~(\ref{beq}), and superimposing this linear $\varepsilon$-$q_{x}$ relationship over the numerically obtained band structures, Fig.~\hyperlink{overlays}{\ref{overlaysplot}}, visually confirms it.

\subsection{Inclusion of long-range hopping parameters}
\label{s:longrange}
\hypertarget{IIc}{}

\begin{figure}
  \hypertarget{disbandsb}{}
  \includegraphics[width=\linewidth]{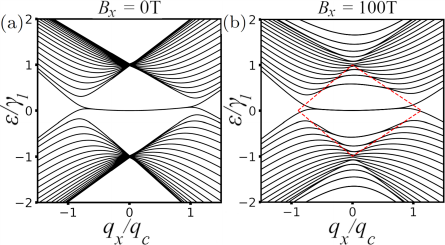}
  \caption{A comparison of band structures for 20 layer rhombohedral graphene with up to next-nearest layer hopping parameters $\gamma_{0,1,2,3,4}$ included, according to the numerical diagonalization of Hamiltonian (\ref{fullhammy}), located at the valley corresponding to $\xi=1$. (a) Shows the band structure for a 20 layer system for $B=0$ to illustrate the difference compared to Fig.~\protect\hyperlink{MAIN2}{1(a)}, (b) shows the same system with the overlay of numerically obtained bifurcation points for $B=100$T. Negative $q_{x}$ values have been included here to highlight anisotropy of the dispersion (trigonal warping) due to the presence of parameter~$\gamma_3$.}
  \label{full HH}
\end{figure}

We consider whether the qualitative features discussed in the context of the minimal model still hold when considering up to next-nearest layer (long-range) hopping parameters in the Hamiltonian.
The full Hamiltonian including these extra parameters $\gamma_{2}$, $\gamma_{3}$, $\gamma_{4}$~\cite{mcclure69,arovas08,koshino09,slizovskiy19} is
\begin{eqnarray}
    H^{N}(\mathbf{q}) =
    \begin{pmatrix}
        \mathcal{A}_{1} & \mathcal{B} &\mathcal{C} &\cdots&0 &0&0 \\
        \mathcal{B}^{\dagger}&\mathcal{A}_{2}&\mathcal{B}&\cdots&0&0&0 \\
        \mathcal{C}^{\dagger} & \mathcal{B}^{\dagger} &\mathcal{A}_{3} & \cdots&0&0&0 \\
        \vdots &\vdots &\vdots & \vdots &\vdots&\vdots&\vdots \\
        0 & 0&0 & \cdots & \mathcal{A}_{N-2} &\mathcal{B}&\mathcal{C}\\
        0&0&0&\cdots& \mathcal{B}^{\dagger} &\mathcal{A}_{N-1} &\mathcal{B}\\
        0&0&0&\cdots& \mathcal{C}^{\dagger} &\mathcal{B}^{\dagger}&\mathcal{A}_{N}
    \end{pmatrix} ,
    \label{fullhammy}
\end{eqnarray}
where each element represents a 2x2 matrix,
\begin{eqnarray*}
    \mathcal{A}_{n} &=&
    \begin{pmatrix}
        0 & \hbar v\pi^{\dagger}_{n}\\
        \hbar v\pi_{n} & 0
    \end{pmatrix} ; \qquad
    \mathcal{B}=
    \begin{pmatrix}
        -\hbar v_{4}\pi^{\dagger} & \hbar v_{3}\pi\\
        \gamma_{1} & -\hbar v_{4}\pi^{\dagger}
    \end{pmatrix} , \\
    \mathcal{C} &=&
    \begin{pmatrix}
        0&\frac{\gamma_{2}}{2}\\
        0& 0
    \end{pmatrix} ; \qquad
v_{i} = \frac{\sqrt{3} a \gamma_{i}}{2\hbar} .
\end{eqnarray*}
The band structure for this Hamiltonian, Fig.~\hyperlink{disbandsb}{\ref{full HH}}, retains significant similarities to the minimal model, but $\gamma_{4}$ introduces finite dispersion with electron-hole asymmetry and $\gamma_{3}$ creates trigonal warping (anisotropy of the dispersion).

For the model~(\ref{hinf}) of infinite layer graphite at zero field, the additional parameters modify matrix elements~\cite{ho13} as
\begin{eqnarray*}
H^{\mathrm{inf}}_{11} &=& H^{\mathrm{inf}}_{22}=2 \hbar v_{4}|\mathbf{q}| \cos (\varphi+k_{z}d) , \\
H^{\mathrm{inf}}_{12} &=& \left(H^{\mathrm{inf}}_{21}\right)^{*} = - \hbar v|\mathbf{q}| e^{-i\varphi}+\gamma_{1}e^{ik_{z}d} \\ 
&& \qquad \qquad \qquad + \, \gamma_{2}e^{-2ik_{z}d}+\hbar v_{3} |\mathbf{q}| e^{i(\varphi-k_{z}d)} ,
\end{eqnarray*}
where $\varphi$ is the polar angle.
Since zero-energy states coincide with a band degeneracy, their location may be determined by setting $H_{12}^{\mathrm{inf}}=0$, and $\gamma_{4}$ has no effect on the location of the nodal line. The solution may be written as a perturbation to Eq.~(\ref{sol}) and, up to $\mathcal{O}(v_{3}/v)$, it can be written as
\begin{eqnarray}
q_{D} &=& \frac{\gamma_{1}}{v\hbar}\left(1+\frac{v_{3}}{v}\cos (3k_{z}d)\right) ,\\
\Phi_{D} &=& -\xi\left(k_{z}d+\frac{v_{3}}{v}\sin (3k_{z}d)\right) ,
\label{newsolu}
\end{eqnarray}
such that the original terms still dominate~\cite{ho13}, and parameter $\gamma_2$ may be neglected.
The presence of $\gamma_3$ (trigonal warping) slightly modifies the semiclassical estimate~(\ref{beq}) of the position of the bifurcations which we determine numerically and show as the dashed overlaid line in Fig.~\hyperlink{disbandsb}{5(b)}.
It has recently been suggested that the sign of $v_3$ is negative~\cite{bao17,garciaruiz23}, in which case one should invert the sign of $q_x$ in Fig.~\hyperlink{disbandsb}{\ref{full HH}(b)}.

\section{Numerics for infinite bulk rhombohedral graphite}
\label{s:infinite}
\hypertarget{III}{}

In Section~\hyperlink{II}{\ref{s:method}}, we described methodology for numerically determining the spectrum of a system with a finite number of layers in an external in-plane field. Here, instead, we describe how to numerically determine the spectrum of bulk rhombohedral graphite (with an infinite number of layers) using a finite-sized magnetic supercell.
Specifically, we consider a magnetic supercell described by Hamiltonian
\small
\begin{eqnarray}
    \mbox{\normalsize $H_{\mathrm{supercell}}^{M}(\mathbf{q})$ }=
    \begin{pmatrix}
        m_{1}&0&0 & \cdots & \Gamma_{1}\\
        0&m_{2}&0 & \cdots & 0 \\
        0&0&m_{3} &\cdots & 0\\
        \vdots&\vdots&\vdots&\vdots& \vdots\\
        \Gamma_{1}^{\dagger}&0&0&\cdots&m_{M}\\
    \end{pmatrix} .
    \label{supercellHamiltonian}
\end{eqnarray}
\normalsize
The magnetic supercell is composed of an integer number $M$ of three-layer hexagonal unit cells, each described by a $6 \times 6$ matrix $m_{n}(\mathbf{q})$ which is similar to the previously considered Hamiltonian~(\ref{HAMMY}),
\small
\begin{multline*}
    \mbox{\normalsize $ m_{n}(\mathbf{q})$ } =\\ \!\!\!\!\!\!
    \begin{pmatrix}
        0 & \hbar v\pi_{3n-2}^{\dagger} & 0 &0 &0 &0\\
        \hbar v\pi_{3n-2} &0 &\gamma_{1}&0&0&0\\
        0&\gamma_{1}&0&\hbar v\pi^{\dagger}_{3n-1}&0&0\\
        0&0&\hbar v\pi_{3n-1}&0&\gamma_{1}&0\\
        0&0&0&\gamma_{1}&0&\hbar v\pi_{3n}^{\dagger}\\
        0&0&0&0&\hbar v\pi_{3n}&0\\
    \end{pmatrix} \! .
\end{multline*}
\normalsize
In the Hamiltonian~(\ref{supercellHamiltonian}), $\Gamma_{1}$ represents a $6\times6$ matrix with $\gamma_1$ as its top right entry and zeros everywhere else, i.e.
$(\Gamma_{1})_{ij} = 0$ for all $i$, $j$, except 
$(\Gamma_{1})_{16} = \gamma_1$. This accounts for periodicity of the supercell.

\begin{figure}
    \centering
    \hypertarget{supercells}{}
    \includegraphics[width=\linewidth]{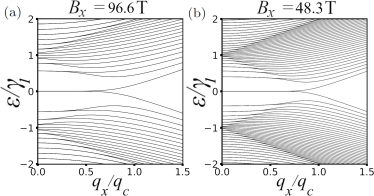}
    \caption{Low-energy band structures for infinite bulk rhombohedral graphene with an applied in-plane magnetic field, obtained through the numerical diagonalization of the supercell Hamiltonian (\ref{supercellHamiltonian}). (a) Shows the band structure for a supercell of size $M=200$, and corresponding field $B_{x}=96.6$T. (b) Is the band structure for a supercell of size $M=400$, and corresponding field $B_{x}=48.3$T.}
\label{supercellsplot}
\end{figure}

We can choose the in-plane component of the magnetic field to lie in the $x$ direction without loss of generality (for the minimal model). It can only take certain discrete values related to the number $M$ of hexagonal unit cells, as determined by periodicity of the Hamiltonian. In particular, for vector potential  ${\bf A}=z(0,-B_x,0)$, a matrix element connecting positions $\mathbf{R}$ and $\mathbf{R^{\prime}}$ has a field-dependent phase factor such as $\Delta = (B_x e/\hbar) \int^{\mathbf{R^{\prime}}}_{\mathbf{R}}z\;dy$. This should be equal to $2\pi$ for a vertical translation of $3Md$ and a horizontal translation of $\sqrt{3}a / 2$, so that allowed values of the magnetic field are
\begin{eqnarray}
B_x = \frac{2h}{3\sqrt{3} adeM} ,
\end{eqnarray}
for integer $M$.
For example, when $M=100$, then $B_x=96.6$T, and the numerical calculation requires the diagonalization of a $600\times600$ matrix~(\ref{supercellHamiltonian}). Despite differences, this method produces similar qualitative results to those for a finite system, section~\hyperlink{II}{\ref{s:method}}, including the location of bifurcations from doubly to singly degenerate states in the energy spectrum as shown in Fig~\hyperlink{supercells}{\ref{supercellsplot}}.

\section{Analogy to the SSH model}
\label{s:ssh}
\hypertarget{IV}{}

For zero magnetic field, the spectrum of RMG in the minimal model may be related to the energy levels of the SSH model~\cite{su79,xiao11,heikkila11,asboth16,cayssol21,mccann23} by dimensional reduction \cite{ryu10}, i.e., by treating the in-plane wave vector $\mathbf{q}$ as a parameter.
In particular, alternating hoppings $t$ and $w$ of the SSH model are equivalent to $\hbar v |{\bf q}|$ and $\gamma_1$, respectively, of RMG.
Here we generalize this analogy in the presence of an in-plane magnetic field, resulting in a one-dimensional tight-binding model with spatially-dependent hopping parameters that bears some similarity to the commensurate off-diagonal Aubry-Andr\'e-Harper model~\cite{ganeshan13}.

Starting from the Hamiltonian~(\ref{HAMMY}) for $N$ layers of RMG, 
we gauge away the phase of the intralayer hopping term for layer $n$ leaving the magnitude,
\begin{eqnarray}
\hbar v \vert\pi_{n}(q_{x},z_{n})\vert = \hbar v r_{n} = \!\sqrt{(\hbar vq_{x})^{2}+(evz_{n}B_{x})^{2}}\,,
    \label{magnitude}
\end{eqnarray}
with $q_y = B_y = 0$. For an even number of layers $N$, the vertical coordinate of the $n$th layer is $z_n = d (n - (N+1)/2)$ with the origin ($z=0$) at the center of the system.

Treating $q_x$ as a parameter, the Hamiltonian~(\ref{HAMMY}) may be written as a one-dimensional tight-binding model which is a generalisation of the SSH model with $2N$ orbitals,
\begin{eqnarray}
    H_{\mathrm{SSH}} = 
    \begin{pmatrix}
        0 & t_{1} & 0 & 0 & 0 & \cdots & 0&0 \\
        t_{1} &  0 & w & 0 & 0 & \cdots & 0&0 \\
        0 & w & 0 & t_{2} & 0 &\cdots & 0 &0\\
        0 & 0 & t_{2} & 0 & w &\cdots & 0 &0\\
        0 & 0 & 0 & w & 0 & \cdots & 0 & 0\\
        \vdots & \vdots & \vdots & \vdots & \vdots & \vdots & \vdots&\vdots \\
        0 & 0 & 0 & 0 & 0 & \cdots & 0 & t_{N}\\
        0&0&0&0&0&\cdots & t_{N}&0\\
    \end{pmatrix}\,, \label{ssh1}
\end{eqnarray}
where
\begin{eqnarray}
t_{n} = \sqrt{t^{2} + \left(\left[n-\tfrac{1}{2}(N+1)\right]\chi\right)^{2}}\,, \label{ssh2}
\end{eqnarray}
and $\chi$ is a parameter describing the strength of the modulated hopping. This has the following equivalence with RMG,
\begin{eqnarray}
t &\equiv& \hbar v |q_x| , \label{equiv1} \\
w &\equiv& \gamma_1 , \label{equiv2} \\
\chi &\equiv& evd |B_x|. \label{equiv3}
\end{eqnarray}
The Hamiltonian~(\ref{ssh1}) has constant intercell hopping $w$ with modulated hopping $t_n$ on every intracell bond.
The modulation is inversion symmetric about the center of the system, increasing in magnitude from $t_{N/2} = \sqrt{t^{2} + (\chi / 2)^{2}} \sim t$ at the center to $t_N \sim \sqrt{t^{2} + (N \chi / 2)^{2}}$ at the edge.

\begin{figure}
    \hypertarget{wf}{}
    \includegraphics[width=\linewidth]{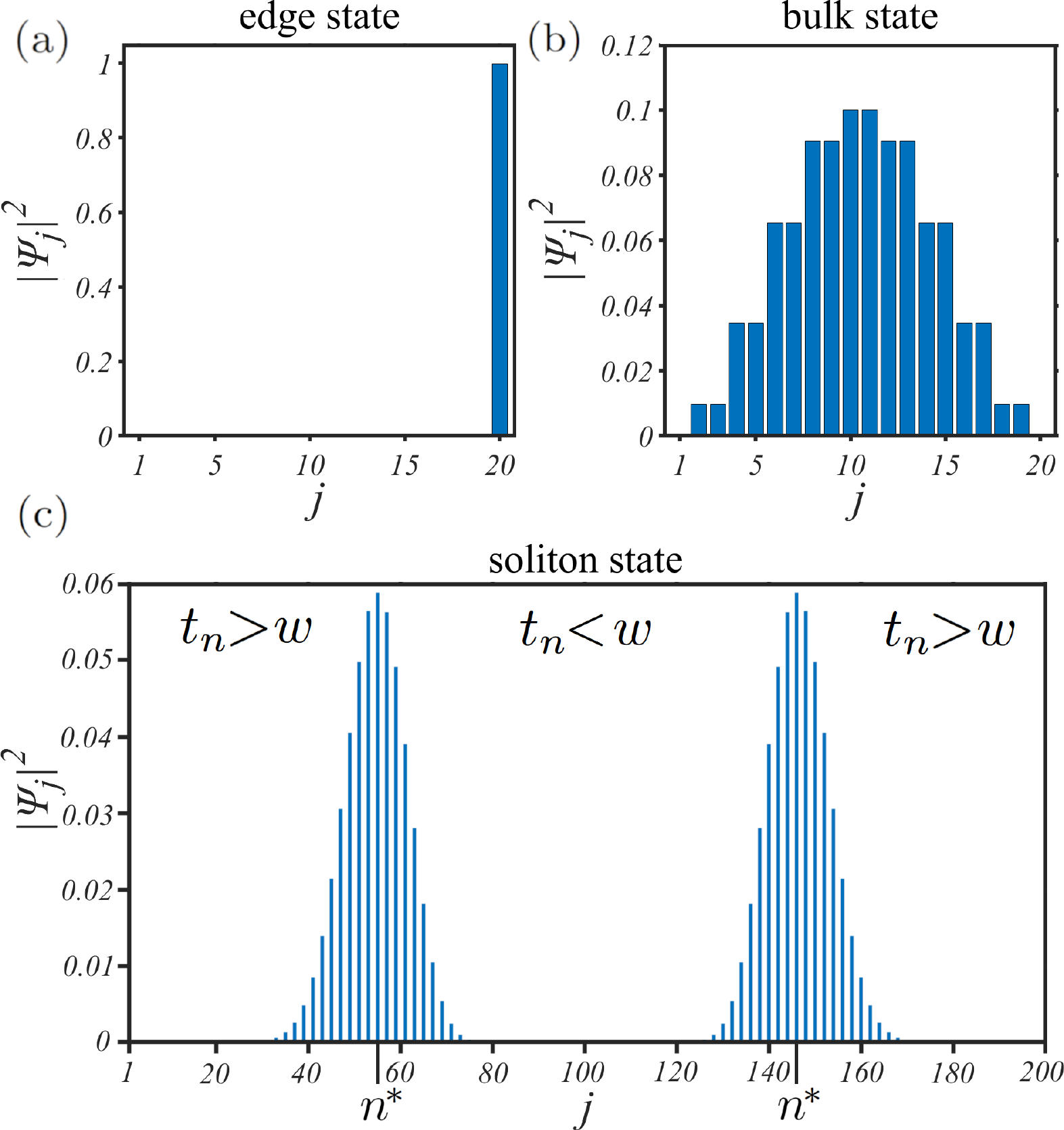} 
    \caption{Probability density of the wavefunctions close to the K-point, $q_{x}=q_{y}=0$, for rhombohedral graphene multilayers on carbon atoms at positions $j = 1,2,3,\ldots,2N$. (a) shows edge state localization at zero energy while (b) is a generic bulk state lying within the valence band. Both (a) and (b) are for a $N = 10$ layer system with $20$ atoms in total and zero magnetic field. (c) shows the distribution of a soliton zero energy state for a $N = 100$ layer system and an in-plane field $B_{x}=50$T. We use a larger layer number to better depict the two soliton states, $n^{\ast}$ denotes the center of the soliton.}
    \label{PROBS}
\end{figure}

We can interpret the effect of $\chi$ as producing a texture of $t$ in the system, i.e. a soliton and anti-soliton pair placed symmetrically about the center, each of which supports a state near zero energy.
For $t < w$ and $\chi = 0$, the Hamiltonian~(\ref{HAMMY}) describes the SSH model with a bulk band gap of $2(w-t)$ and states near zero energy localized at the edges, Fig.~\hyperlink{wf}{\ref{PROBS}(a)}. By contrast, Fig.~\hyperlink{wf}{\ref{PROBS}(b)} shows the spatial distribution of a state lying within the bulk valence band.
For nonzero $\chi$, and large enough $N$, $t_N$ at the edge becomes larger than $w$ so that there are no edge states. However, at some position $n=n^{\ast}$  there will be $t_{n^{\ast}} \approx w$ separating regions with $t_n > w$ towards the edge and with $t_n < w$ towards the center. Location $n^{\ast}$ is at the center of a soliton and supports a zero energy state, and there is a corresponding antisoliton with the opposite texture of $t_n$ at the other side of the system which also supports a zero energy state, Fig.~\hyperlink{wf}{\ref{PROBS}(c)}. Thus, as $\chi$ increases, the soliton-antisoliton pair move from the edges towards the center of the system, always supporting zero energy states. Finally, for large enough $\chi$, they will reach the center and annihilate, when
$t_{N/2} = \sqrt{t^{2} + (\chi / 2)^{2}} \approx w$, and there will be no zero energy states.

\begin{figure}
\centering
\hypertarget{zandq}{}
\includegraphics[width=0.8\linewidth]{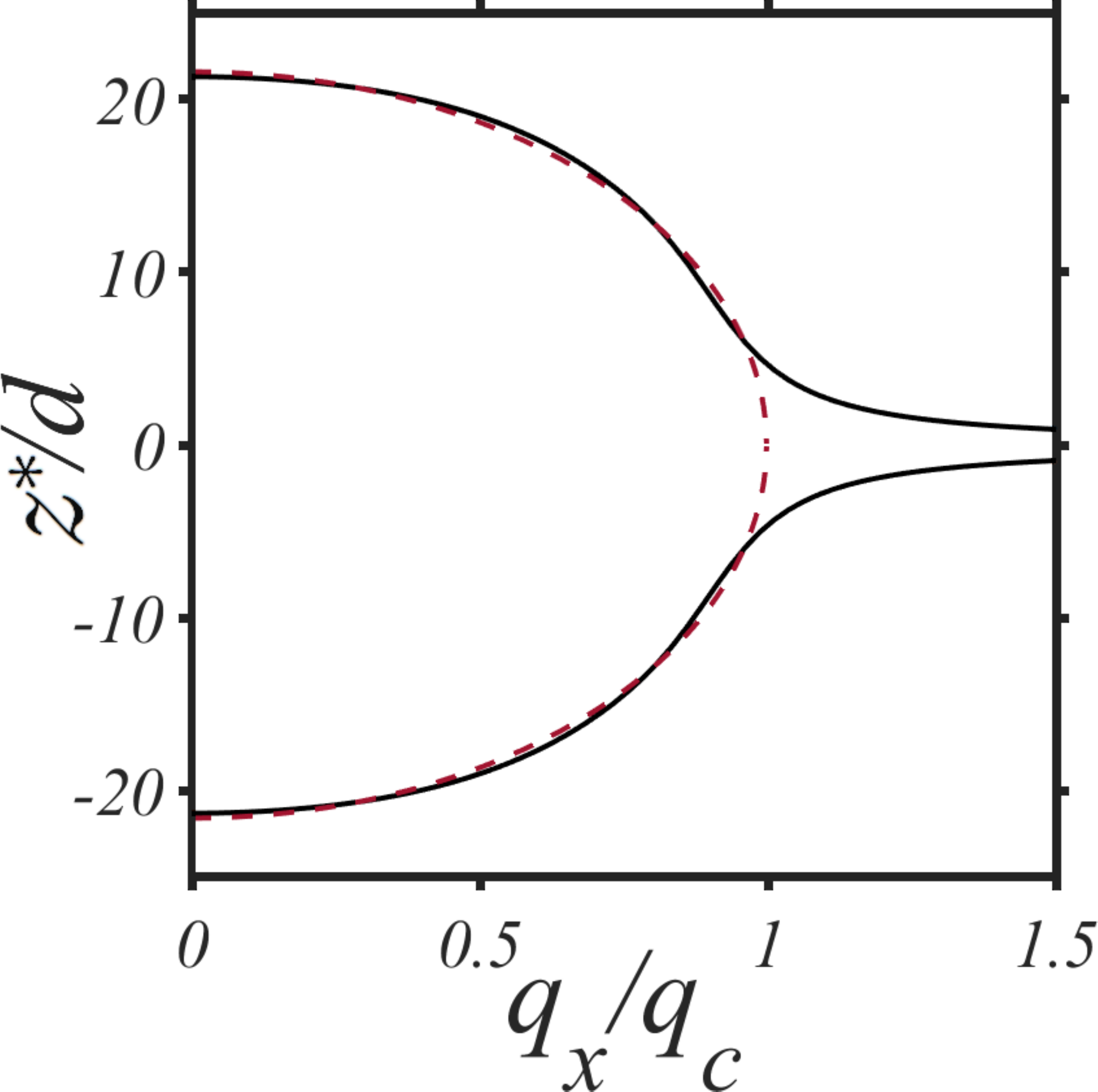}
\caption{Vertical position $z^{\ast}$, as a function of the in-plane wave vector $q_{x}$, of the localized states with zero-energy flat bands in rhombohedral multilayer graphene with $N= 50$ layers and an applied in-plane magnetic field $B_{x}=50$T. Solid lines are obtained by numerically determining the peak positions of the wavefunctions, the dashed line is the estimate Eq.~(\ref{zast}).
States at $q_{x}=0$ are localized at high $|z|$ values towards the surface of the thin film. As $q_{x}$ increases, they move inwards to the center at which point they begin to annihilate one another, asymptotically approaching the minimum of $|z|=d/2$.}
\label{zandqplot}
\end{figure}

For fixed $\chi$, the location of the soliton $n^{\ast}$ is dependent on the value of $t$. From $t_{n^{\ast}} \approx w$ it may be estimated at the right side of the system ($n^{\ast} > (N+1)/2$), say, to be
$n^{\ast} \approx \mathrm{min} \big( N ,  (N+1)/2 + \chi^{-1} \sqrt{w^2 - t^2} \big)$.
Using the equivalence to RMG, Eqs.~(\ref{equiv1}-\ref{equiv3}), we can express the layer position $z^{\ast} \equiv \big( n^{\ast} - (N+1)/2 \big) d$ where the flat band state is localized as
\begin{eqnarray}
\frac{z^{\ast}}{d} \approx \mathrm{min} \bigg( \frac{N-1}{2} ,    \frac{a \sqrt{q_c^2 - |q_x|^2}}{2\pi (\phi_d / \phi_0)} \bigg) ,
\label{zast}
\end{eqnarray}
for $|q_x| < q_c$ where $q_c = \gamma_1 / (\hbar v)$,
$\phi_d = ad|B_x|$ is the magnetic flux per unit cell,
and $\phi_0 = h/e$ is the single particle flux quantum.
This expression is for the top half of the thin film ($z^{\ast} > 0$), and the position for the lower half is $-z^{\ast}$.
Fig~\hyperlink{zandq}{\ref{zandqplot}} shows a plot of $z^{\ast}$ obtained by numerically determining the peak positions of the wavefunctions, with the estimate Eq.~(\ref{zast}) shown for comparison as the dashed line.
It shows that the zero-energy flat bands are localized on different bulk layers of the system, not just the surfaces, as determined by the wave vector $q_x$.
This accounts for the behavior observed for the flat bands at zero energy in RMG in the presence of an in-plane magnetic field.
As the field strength is increased, the zero energy states persist, but the extent of the flat bands in $q_x$ is reduced because, for large $q_x$ values, the solitons have disappeared. Finally, at very large field strengths, they separate and move into the bulk (when the solitons for all $q_x$ values have vanished).

By equating the two terms on the right side of Eq.~(\ref{zast}), we can estimate the extent of the flat band, in terms of $q_x$, that remains on the surface in the presence of the in-plane field as
\begin{eqnarray}
q_x a = \sqrt{ q_c^2 a^2 - (\pi \phi_N / \phi_0)^2} ,
\end{eqnarray}
where the magnetic flux summed over all layers is $\phi_N = (N-1)\phi_d = ad (N-1) B_x$.

\begin{figure}
    \centering
    \hypertarget{loglog}{}
    \includegraphics[width=0.97\linewidth]{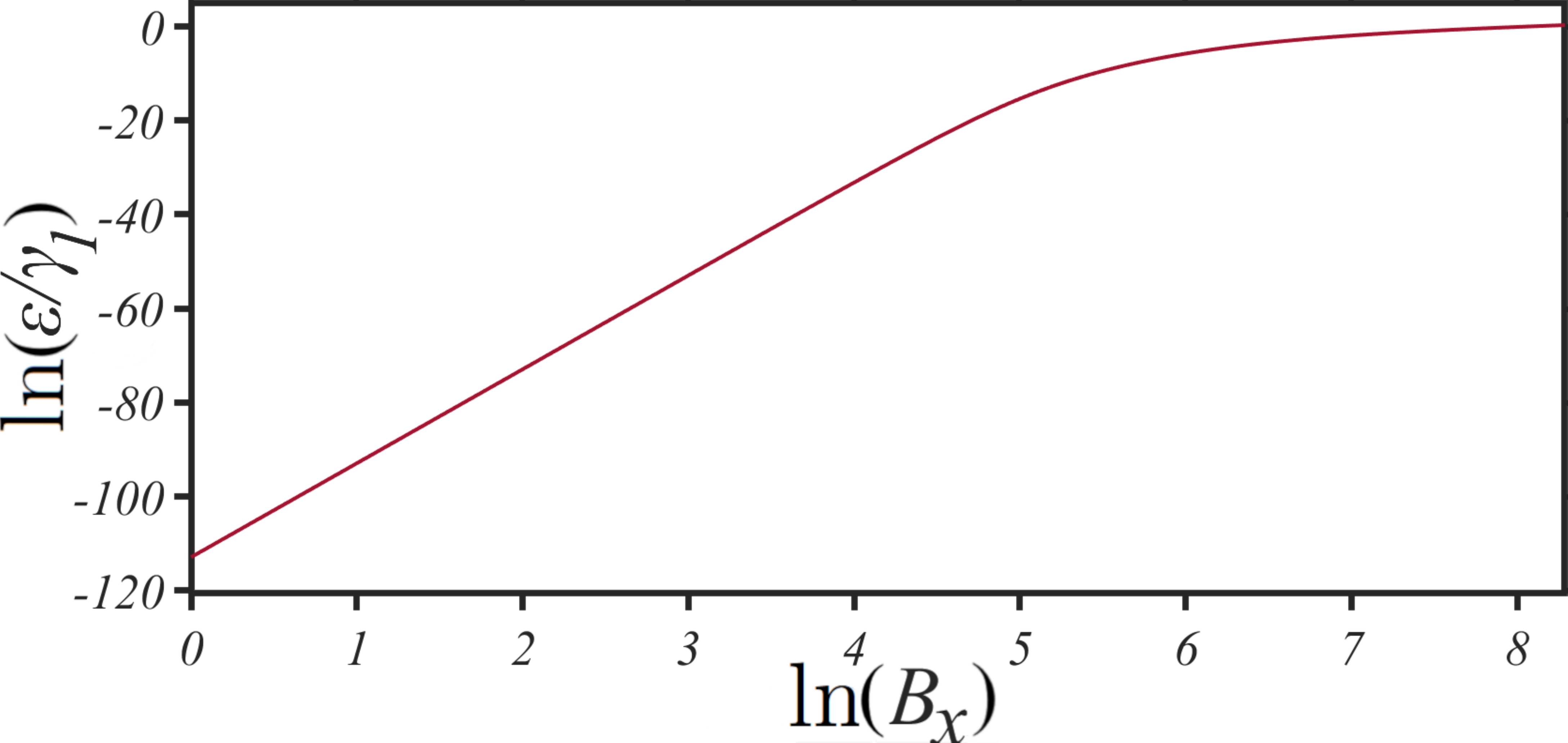}
    \caption{Energy of the states near zero energy at $q_{x}=0$ in $N=20$ layer rhombohedral graphene as a function of magnetic field $B_x$, represented as a log-log plot. States localized on surfaces of the RMG film at small magnetic field are characterized by a linear relation between energy and magnetic field in the log-log plot, reflecting the analytical solution $\varepsilon\propto B_{x}^{N}$.}
\label{loglogplot}
\end{figure}

The limit $t=0$ for the SSH model corresponds to $q_x = 0$ in RMG, in which case $t_N \sim N \chi /2$ at the edge. Thus, the zero energy states for $t=0$ remain at the edge until $t_N \approx w$, i.e., $\chi \sim 2w/N$. In terms of RMG, this means that states at $q_x = 0$ remain on the surfaces until the magnetic flux $\phi_N$ reaches values
\begin{eqnarray}
\frac{\phi_N}{\phi_0} \sim \frac{2\gamma_1}{\sqrt{3} \pi \gamma_0} ,
\label{est1}
\end{eqnarray}
where $\phi_0 = h/e$ is the single particle flux quantum; numerical values give $\phi_N \sim \phi_0 /20$.
We can alternatively write this estimate as $\phi_N/\phi_0 \sim q_c a / \pi$.
Up to this point, we can use the zero field estimate for the energy of the surface states, $\varepsilon \approx \pm (\hbar v |{\bf q}|)^N / \gamma_1^{N-1}$, with the substitution $|{\bf q}| \rightarrow B_x$ to estimate that the surface states at $q_x = 0$ behave as $\varepsilon \sim \pm B_x^N$.
This is confirmed numerically in Fig.~\hyperlink{loglog}{\ref{loglogplot}}, which also depicts larger magnetic fields breaking this relationship as the localized states move from the edges towards the center.
The localized states will eventually be destroyed when
$t_{N/2} = \sqrt{t^{2} + (\chi / 2)^{2}} \approx w$ and, for $t=0$, this gives an estimate $\chi \sim 2w$. 
In terms of RMG, this means that the zero energy states remain until
\begin{eqnarray}
\frac{\phi_N}{\phi_0} \sim \frac{2N\gamma_1}{\sqrt{3} \pi \gamma_0} ,
\end{eqnarray}
We can alternatively write this estimate as $\phi_d/\phi_0 \sim q_c a / \pi$ where $\phi_d = ad|B_x|$ is the magnetic flux per unit cell.
Clearly, this is a huge magnetic field, a large factor (of $N \gg 1$) greater than the field~(\ref{est1}) for which the zero energy states move away from the edge.

We determine the density of states per unit energy per unit area (DOS) to show how the flat bands at zero energy contribute to sharp features in the DOS. Since the Hamiltonian~(\ref{HAMMY}) satisfies sublattice chiral symmetry in the same way as the SSH model, we check that the zero energy states are robust to chiral-preserving disorder by calculating the disorder-averaged DOS in the presence of disorder.
We calculate the DOS per unit energy numerically by approximating the Dirac delta function as a Lorentzian with finite width $\zeta = 0.005\gamma_{1}$,
\begin{eqnarray}
    g(\varepsilon) = \frac{1}{\pi L^2}\sum_{n}\frac{\zeta}{(\varepsilon-\varepsilon_{n})^{2}+\zeta^{2}} ,
    \label{DOSSS}
\end{eqnarray}
where $L^2$ is the area of the system.

\begin{figure}
    \hypertarget{DOSS}{}
    \includegraphics[width=\linewidth]{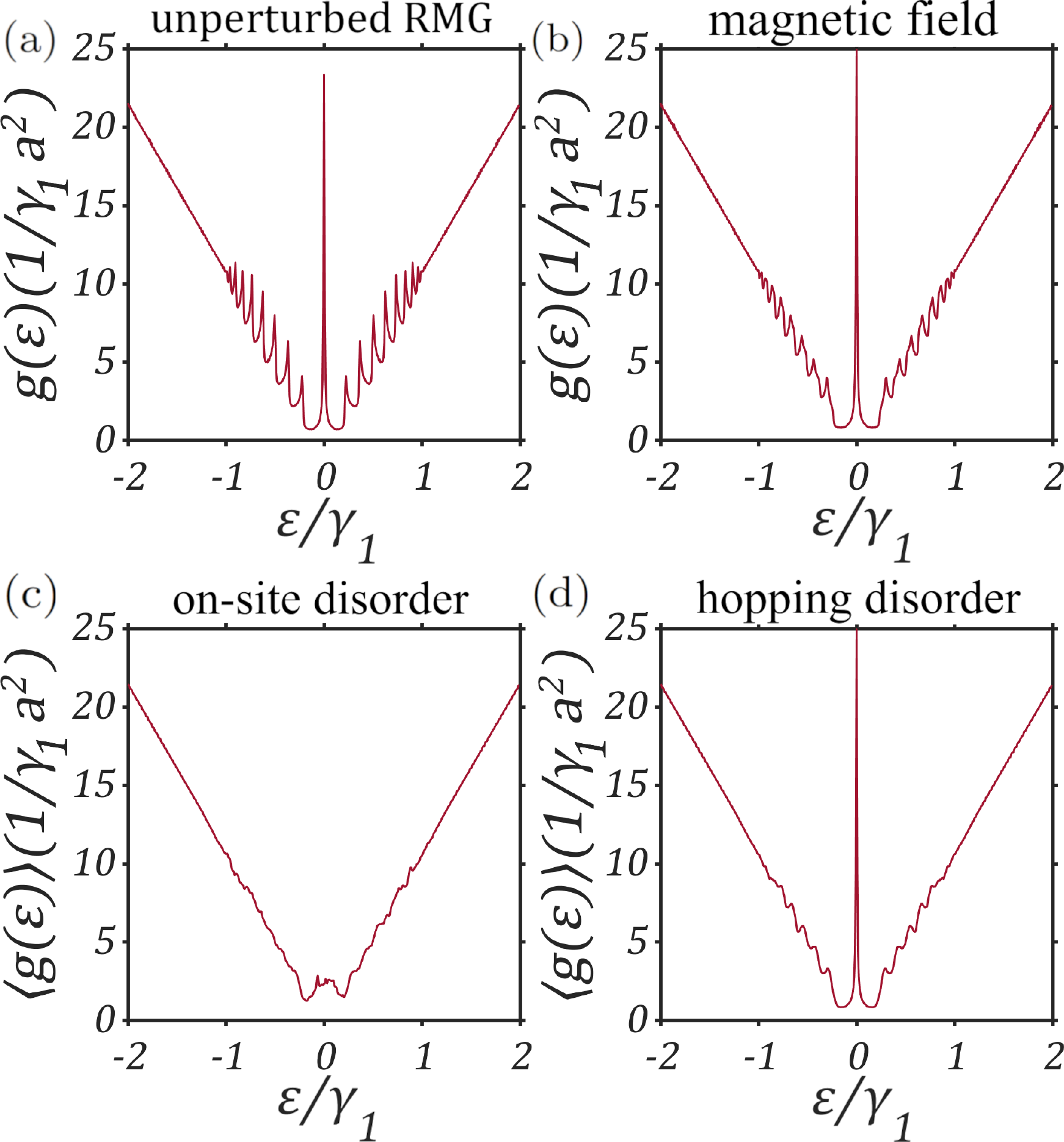}
    \caption{The DOS as a function of energy for $N=20$ layer  rhombohedral graphene systems calculated numerically using the minimal model Hamiltonian~(\ref{HAMMY}) and DOS definition~(\ref{DOSSS}). (a) depicts DOS for pristine RMG at zero field. (b) pristine RMG with an applied in-plane magnetic field of $B_x = 50\,$T. (c) is the disorder-averaged DOS at $B_x = 50\,$T created by averaging over 20 randomized disorder realizations for chirality-destroying disorder. (d) disorder-averaged DOS at $B_x = 50\,$T for 20 realizations of chirality-preserving disorder.}
    \label{DOSSplot}
\end{figure}

For pristine RMG at zero magnetic field, the DOS is characterized by a series of Van Hove singularities at energies corresponding to turning points of the bulk bands, Fig.~\hyperlink{DOSS}{\ref{DOSSplot}a}, with a prominent central peak at zero energy caused by the flat bands.
The DOS in a magnetic field of $B_{x}=50$T is shown in Fig.~\hyperlink{DOSS}{\ref{DOSSplot}b}, and is characterized by the softening of Van Hove singularities, but the peak corresponding to flat bands remains as prominent as before.

We introduce chirality-breaking disorder through the addition of onsite energies randomly chosen from a distribution $[-\delta , \delta]$, where $\delta=50\mbox{meV}\ll\gamma_{1}$~\cite{muten21}.
The resulting disorder-averaged DOS in the presence of a magnetic field is shown in Fig.~\hyperlink{DOSS}{\ref{DOSSplot}c} with averaging over 20 disorder realisations.
We also consider chirality-preserving disorder in the form of random additions $\delta$ to the interlayer coupling $\gamma_{1}$ as $\gamma_{1} + \delta$,
chosen from a distribution $[-\delta , \delta]$, where $\delta=50\mbox{meV}\ll\gamma_{1}$. The result of averaging over 20 disorder realisations in a finite magnetic field is shown in Fig.~\hyperlink{DOSS}{\ref{DOSSplot}d}.

The DOS for both types of disorder are characterized by further softening of the Van Hove singularities at finite energies. However, chirality-destroying disorder destroys the zero energy flat band peak, whereas chirality-preserving disorder has no visible effect on the zero-energy peak.
This is expected because the zero energy states rely on the protection of chiral symmetry~\cite{inui94,asboth16,scollon20,bellec20}.
In the plot of Fig.~\hyperlink{DOSS}{\ref{DOSSplot}c}, for chirality-destroying disorder, there is also electron-hole asymmetry in the DOS, but this is only an artefact of having a finite number of disorder realisations.

\section{Conclusion}
\label{s:conclusion}
\hypertarget{V}{}

We numerically determine the effect of an in-plane magnetic field on the electronic spectrum of rhombohedral multilayer graphene (RMG).
The spectrum shows two qualitative features: Firstly, there are a series of bifurcations from doubly- to singly-degenerate energy levels and, secondly, the zero-energy flat bands, observed at zero field, persist up to very high field strengths.
Using semiclassical quantization~\cite{onsager52,kormanyos08,fuchs10,xiao10} of the zero field Fermi surface of rhombohedral graphite, we are able to relate the presence of level bifurcations to  bifurcations of zero field isoenergetic contours from one contour, with zero Berry phase, to two contours each with Berry phase $\pi$.
By writing down a one-dimensional tight-binding model that is analogous to RMG in an in-plane field, we are able to relate the persistence of the zero-energy bands in large magnetic fields to a soliton texture supporting zero-energy states in the Su-Schrieffer-Heeger model.
Whereas the zero-energy flat bands at zero field are localized on the surfaces of the RMG thin film, in a finite magnetic field they are generally localized on different layers, ranging from the surface to the bulk as a function of the wave vector. 

The electronic properties of RMG films may be experimentally accessed by a variety of methods including scanning tunneling spectroscopy~\cite{pierucci15}, magneto-Raman~\cite{faugeras09,yan10,faugeras11,kuhne12,goler12,qiu13,henni16},  photoemission~\cite{pierucci15,henck18}, and magneto-transport~\cite{chiappini16,yin19,shi20,mullan23}.
For example, scanning tunneling microscopy was recently used to probe the surface states of RMG with up to $N = 17$ layers~\cite{hagymasi22}, and magneto-transport properties, probing the Landau level spectra, have been measured in RMG with up to $N = 50$ layers~\cite{shi20} and in Bernal-stacked graphitic films with up to a few hundred layers~\cite{yin19,mullan23}.

We neglected spin splitting and assumed a fourfold degeneracy due to spins and valleys  throughout. In an external in-plane magnetic field, splin splitting of the energy levels will occur.
With $g$-factor $g=2$, the spin splitting should be $\Delta \varepsilon = 2 \mu_B B$ for Bohr magneton $\mu_B$ and field strength $B$. Thus, for $B=100\,$T, we expect $\Delta \varepsilon = 0.012\,$eV and $\Delta \varepsilon / \gamma_1 = 0.030$.
Experiments~\cite{kurganova11} in monolayer and bilayer graphene reported an effective $g$-factor, $g^{\ast} = 2.7 \pm 0.2$, and attributed its enhancement to electron-electron interaction effects. Nevertheless, even with such an enhancement, we expect the spin splitting to be a small effect on the scale of our typical plots, e.g. Fig.~\hyperlink{MAIN2}{\ref{MAIN}(b)}.

All relevant data presented in this paper can be accessed~\cite{datanote}.

\begin{acknowledgments}
The authors thank A. Mishchenko and S. Ozdemir for helpful discussions.
\end{acknowledgments}


\begin{thebibliography}{99}
\bibitem{magnitudesnote}
For clarity of presentation, we show numerical data for systems with a relatively small number of layers and very large magnetic fields, see Fig.~\hyperlink{MAIN2}{1(b)}. However, similar qualitative features appear for lower magnetic field strengths in samples with more layers, Fig~\hyperlink{largeN}{\ref{LargeNBands}}, although bifurcations in the energy spectrum are less visible for these parameter values.
\end{thebibliography}
\end{document}